\documentclass[preprint,showpacs,preprintnumbers,amsmath,amssymb]{revtex4}

\usepackage{graphicx}% Include figure files
\usepackage{dcolumn}% Align table columns on decimal point
\usepackage{bm}% bold math

\begin{document}

%\preprint{APS/123-QED}

\title{Magnetic order in Pr$_{2}$Pd$_{3}$Ge$_{5}$ and possible heavy fermion behavior in Pr$_{2}$Rh$_{3}$Ge$_{5}$}
\author{V. K. Anand}
\email{vivekkranand@gmail.com}
\author{Z. Hossain}
%\email{zakir@iitk.ac.in}
\affiliation{ Department of Physics, Indian Institute of Technology, Kanpur 208016, India}
\author{C. Geibel}
\affiliation{Max Planck Institute for Chemical Physics of Solids, 01187 Dresden, Germany}
%\email{ }
%\homepage{}
\date{\today}% It is always \today, today,
\begin{abstract}
         We report our results on two new ternary intermetallic compounds Pr$_{2}$Pd$_{3}$Ge$_{5}$ and Pr$_{2}$Rh$_{3}$Ge$_{5}$ based on magnetic susceptibility, magnetization, specific heat, resistivity and magnetoresistance data.  These compounds form in U$_{2}$Co$_{3}$Si$_{5}$-type orthorhombic structure (space group \textit{Ibam}).  Pr$_{2}$Pd$_{3}$Ge$_{5}$ exhibits two magnetic transitions at T$_{N1}$ = 7.5 K and T$_{N2}$ = 8.3 K. In the magnetically ordered state Pr$_{2}$Pd$_{3}$Ge$_{5}$ exhibits a field induced metamagnetic transition and unusually large magnetoresistance. Pr$_{2}$Rh$_{3}$Ge$_{5 }$ does not show any phase transition down to 0.5 K. It has a CEF singlet ground state, separated from the first excited state by about 10 K. The low lying crystal field excitations lead to exciton mediated electronic mass enhancement as evidenced by large Sommerfeld coefficient ($\gamma$ $\sim$ 80 mJ/mol K$^2$) in Pr$_{2}$Rh$_{3}$Ge$_{5}$.
\end{abstract}

\maketitle

%\section{\label{}Introduction\protect \lowercase{} }

\section*{Introduction}

         With the advent of unconventional heavy fermion superconductivity in PrOs$_{4}$Sb$_{12}$,  the Pr-based compounds have evolved as a topic of current interest in condensed matter physics \cite{1,2,3,4}. Ce-compounds present many examples of Kondo lattice/heavy fermion behaviour and magnetically mediated unconventional superconductivity which occurs in the magnetic-nonmagnetic boundary. In contrast, so far we know of only one example of Pr-based heavy fermion superconductor. The physics of heavy fermion behaviour in Pr-compounds is very different from that in Ce-compounds. While in Ce-compounds the heavy fermion behaviour has its origin in Kondo effect, in Pr-compounds it is realized either due to the quadrupolar Kondo effect as in PrInAg$_{2}$ \cite{5} or due to the excitonic mass enhancement as observed in PrOs$_{4}$Sb$_{12}$ \cite{6}. Both the routes of heavy fermion behaviour in Pr-compounds critically depend on the crystal electric field (CEF) split ground state. The magnetic and transport properties of Pr-compounds are strongly influenced by the crystal electric field effect. Exciting phenomena are observed in the Pr-based systems with nonmagnetic singlet/doublet ground state when the crystal field splitting energy becomes comparable with other interactions. Realizing that the Pr-based compounds can also present extraordinary magnetic and superconducting properties we have started working on these systems with an objective of finding novel Pr-compounds and explore the interplay of magnetism and superconductivity in them.  In order to achieve the destined goal we are investigating Pr$_{2}$T$_{3}$Ge$_{5}$ compounds (T = transition elements) in view of interesting features of their Ce-analogs. As for example, Ce$_{2}$Ni$_{3}$Ge$_{5}$, which is a Kondo lattice antiferromagnetic system, exhibits pressure induced superconductivity around 3.9 GPa \cite{7,8}. Very recently we investigated its Pr-analog, Pr$_{2}$Ni$_{3}$Ge$_{5}$, and found an antiferromagnetic ordering below 8.5 K and two field-induced metamagnetic transitions as well as very high positive magnetoresistance in the ordered state \cite{9}. In this article we report our findings on magnetic and transport properties of Pr$_{2}$Pd$_{3}$Ge$_{5}$ and Pr$_{2}$Rh$_{3}$Ge$_{5}$. Ce-analogs of these two compounds viz. Ce$_{2}$Pd$_{3}$Ge$_{5}$ orders antiferromagnetically at 3.8 K \cite{10}, and Ce$_{2}$Rh$_{3}$Ge$_{5}$ is a moderate heavy-fermion antiferromagnet system \cite{11}. The homologous compound Pr$_{2}$Rh$_{3}$Si$_{5}$ does not order magnetically down to 1.8 K due to the nonmagnetic singlet CEF ground state \cite{12}.

\section*{Experimental}

          We prepared polycrystalline samples of Pr$_{2}$Pd$_{3}$Ge$_{5}$ and Pr$_{2}$Rh$_{3}$Ge$_{5 }$ and the nonmagnetic La-analogs starting with high purity elements in stoichiometric ratio by the conventional arc melting on a water cooled copper hearth under argon atmosphere. During the arc melting process samples were flipped and remelted several times to improve the homogeneity. Arc melted samples were annealed at 1000$^{o}$C under vacuum for 7 days. Both annealed and as-cast samples were characterized by Copper K$_\alpha$ x-ray diffraction and scanning electron microscopy equipped with energy dispersive x-ray analysis (EDAX). Magnetization was measured using a commercial SQUID magnetometer. Resistivity was measured using standard four probe ac technique. Heat capacity was measured using relaxation method in a physical property measurement system (PPMS-Quantum design). A second batch of Pr-samples was also prepared to check for the reproducibility of the results.

\section*{Results and discussion}

           X-ray diffraction data on powdered samples of R$_{2}$T$_{3}$Ge$_{5}$ (R = Pr, La; and T = Pd, Rh) were analyzed using the software WINXPOW. These compounds crystallize in U$_{2}$Co$_{3}$Si$_{5}$-type orthorhombic structure (space-group \textit{Ibam}). The lattice parameters and unit-cell volumes are listed in Table I. Pr$_{2}$Pd$_{3}$Ge$_{5}$ and Pr$_{2}$Rh$_{3}$Ge$_{5}$ have slightly lower lattice volumes compared to their Ce-analogs, and the lattice parameters of La-compounds are in agreement with the reported values \cite{10,11}. Powder x-ray diffraction and scanning electron micrographs revealed the samples to be essentially single phase (impurity phase(s) less than 3\%). EDAX composition analysis revealed the samples to have the expected 2:3:5 compositions.

\begin{table}
\caption{\label{tab:table1}Lattice parameters and unit cell volumes of orthorhombic (\textit{Ibam}) system R$_{2}$T$_{3}$Ge$_{5}$ (R = Pr, La; and T = Pd, Rh).}
\begin{ruledtabular}
\begin{tabular}{lcccc}
 compounds &$a$ (\AA)&$b$ (\AA)&$c$ (\AA) &$V$ (\AA$^3$)\\
\hline
Pr$_{2}$Pd$_{3}$Ge$_{5}$& 10.150 & 12.084 & 6.144 & 753.6 \\
La$_{2}$Pd$_{3}$Ge$_{5}$& 10.186 & 12.223 & 6.182 & 769.6 \\
Pr$_{2}$Rh$_{3}$Ge$_{5}$& 10.078 & 12.091 & 5.978 & 728.4 \\
La$_{2}$Rh$_{3}$Ge$_{5}$& 10.149 & 12.185 & 6.033 & 746.1 \\
\end{tabular}
\end{ruledtabular}
\end{table}

\section*{A. Pr$_{2}$Pd$_{3}$Ge$_{5}$}

          Figure 1 shows the temperature dependence of magnetic susceptibility of Pr$_{2}$Pd$_{3}$Ge$_{5}$ at 1.0~T. The low temperature susceptibility data at different fields are shown in the inset of figure 1. At low fields (0.01~T) two well pronounced transitions are observed in the magnetic susceptibility data at 8.3~K and 7.5~K. On increasing the strength of the applied field both anomalies merge together and the position of peak shifts to lower temperature (e.g. to 7.2~K at 1.5~T) evidencing the onset of an antiferromagnetic ordering at 8.3 K. At a field of 3 T and above, the susceptibility curve shows a tendency to saturation. The magnetic susceptibility follows the modified Curie-Weiss behavior $\chi$ = $\chi_{0}$ + C/(T-$\theta_{p}$). At 1~T the susceptibility data above 50 K fits nicely with $\chi_{0}$ = -8 x10$^{-5}$ emu/Pr-mole and the paramagnetic Curie-Weiss temperature $\theta_{p}$ = -6.1 K. The effective moment $\mu_{eff }$= 3.53 $\mu_{B}$ is very close to the theoretically expected value for Pr$^{3+}$ ions (3.58 $\mu_{B}$).

\begin{figure}
\includegraphics[width=8.5cm, keepaspectratio]{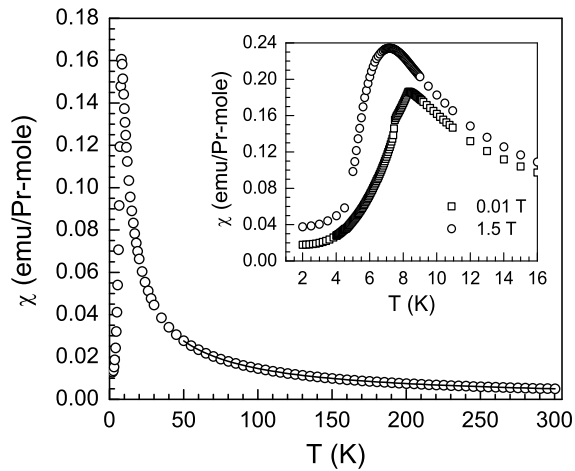}
\caption{\label{fig1} The temperature dependence of magnetic susceptibility of Pr$_{2}$Pd$_{3}$Ge$_{5}$ at 1.0 T. Solid line represents the fit to the modified Curie-Weiss behaviour. Inset shows the low temperature magnetic susceptibility data at two different fields.}
\end{figure}

\begin{figure}
\includegraphics[width=8.5cm,keepaspectratio]{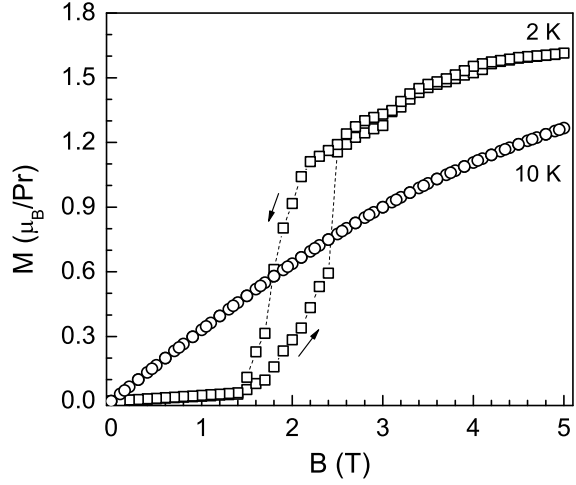}
\caption{\label{fig2} Isothermal magnetization of Pr$_{2}$Pd$_{3}$Ge$_{5}$ as a function of field at temperatures 2 K and 10 K.}
\end{figure}

\begin{figure}
\includegraphics[width=8.5cm,keepaspectratio]{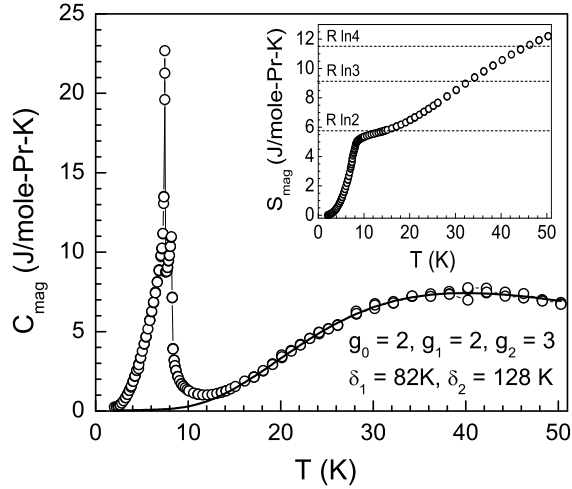}
\caption{\label{fig3} Magnetic part of the specific heat of Pr$_{2}$Pd$_{3}$Ge$_{5}$. Solid line represents the fit to the three level CEF scheme as described in the text.  Inset shows the magnetic contribution to the entropy as a function of temperature.}
\end{figure}

         From the isothermal magnetization data of Pr$_{2}$Pd$_{3}$Ge$_{5}$ shown in figure 2 we observe a rapid increase in magnetization at 1.6 T at low temperatures (2K which is well below T$_{N}$) due to the occurrence of field induced metamagnetic transition. This field induced transition is accompanied by a large hysteresis. A slight nonlinearity in magnetization curve is found at 10 K which is just above the magnetic ordering temperature. This may result from the presence of short range order above the magnetic ordering temperature and/or due to a contribution from the crystal electric field effect. At higher temperatures magnetization exhibits a linear field dependence. In the ordered state saturation magnetization is 1.65 $\mu_{B}$/Pr at 5T.

         Figure 3 shows the magnetic contribution to the specific heat of Pr$_{2}$Pd$_{3}$Ge$_{5}$ which we estimated by subtracting the specific heat of the nonmagnetic compound La$_{2}$Pd$_{3}$Ge$_{5}$ from the specific heat of Pr$_{2}$Pd$_{3}$Ge$_{5}$, assuming the lattice contribution to be roughly equal to that of La$_{2}$Pd$_{3}$Ge$_{5}$. The specific heat data of La$_{2}$Pd$_{3}$Ge$_{5}$ do not show any anomaly down to 2 K and has a $\gamma$ value of $\sim$ 5 mJ/mole-La-K$^{2}$ (below 10 K). The specific heat data of Pr$_{2}$Pd$_{3}$Ge$_{5}$ exhibit two sharp $\lambda$-type anomalies at 8.2 K and 7.5 K, confirming the intrinsic nature of magnetic ordering observed in magnetic susceptibility data. The anomaly at 7.5 K in specific heat is more pronounced in contrast to that in magnetic susceptibility. The specific heat data fits very nicely to the expresssion C = $\gamma$T + $\beta$T$^{3}$exp(-E$_{g}$/k$_{B}$T) in the temperature range (2--7 K). This clearly means that the magnon spectrum has an energy gap in antiferromagnetic state which in turn suggests the possibility of anisotropic magnetic behaviour in this compound. The coefficients $\gamma$ = 14 mJ/mole-Pr-K$^{2}$ and $\beta$ = 0.059 J/mol-Pr-K$^{4}$, and the energy gap E$_{g}$ = 0.25 meV.

          At higher temperatures the magnetic part of specific heat shows a Schottky-type anomaly in the form of a broad peak centered around 40 K. The experimentally observed magnetic specific heat data above 20 K could be reproduced by the crystal field scheme involving three levels: a doublet ground state separated by 82 K from a doublet first excited state and 128 K from the triplet second excited state. The solid line in figure 3 represents this CEF scheme and includes an additional electronic contribution with $\gamma$ $\sim$ 10 mJ/Pr-mole-K$^{2}$ (over that of La$_{2}$Pd$_{3}$Ge$_{5}$ which we subtracted while estimating the magnetic contribution to the specific heat). The magnetic contribution to the entropy attains a value of 5.35 J/mole-Pr-K at 10 K which is 93\% of Rln(2). 

\begin{figure}
\includegraphics[width=8.5cm,keepaspectratio]{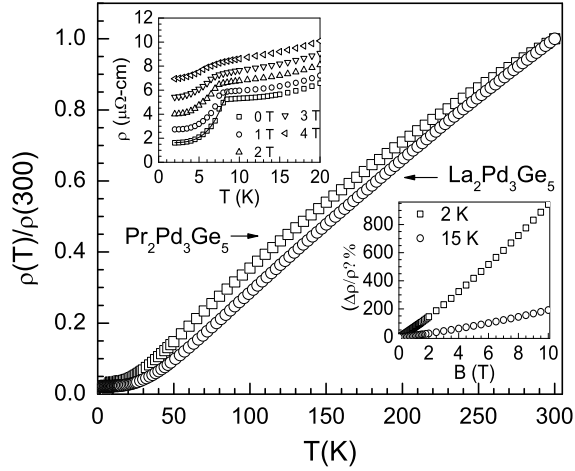}
\caption{\label{fig4} Normalized electrical resistivity of La$_{2}$Pd$_{3}$Ge$_{5}$ and Pr$_{2}$Pd$_{3}$Ge$_{5}$ as a function of temperature in the temperature range of 2 K -- 300 K. The upper inset shows the low temperature resistivity of Pr$_{2}$Pd$_{3}$Ge$_{5}$ at different fields, and the lower inset shows the magnetoresistance at 2 K and 15 K.  The solid line in the upper inset represents the fit to gapped magnon behaviour in ordered state.}
\end{figure}

          The results on electrical resistivity measurements of La$_{2}$Pd$_{3}$Ge$_{5}$ and Pr$_{2}$Pd$_{3}$Ge$_{5}$ are shown in figure 4. The resistivity data of La$_{2}$Pd$_{3}$Ge$_{5}$ have almost linear temperature dependence with a residual resistivity of $\sim$ 4.5 $\mu\Omega$-cm at 2 K and residual resistivity ratio $\sim$ 44. In the case of Pr$_{2}$Pd$_{3}$Ge$_{5}$, residual resistivity at 2 K is $\sim$ 1.6 $\mu\Omega$-cm and residual resistivity ratio $\sim$ 90. The low residual resistivities and high residual resistivity ratios indicate our polycrystalline samples to be of very good quality. The resistivity data of Pr$_{2}$Pd$_{3}$Ge$_{5}$ exhibit slight nonlinearity which we attribute to crystal field effect. At low temperatures (below 15 K) the resistivity attains a nearly constant value before it undergoes a transition at T$_{N}$ below which there is a sharp drop of resistivity due to reduction in spin disorder scattering. The low temperature resistivity data of Pr$_{2}$Pd$_{3}$Ge$_{5}$ at different magnetic fields up to 4 T are shown as inset in figure 4. With the increase in field strength the resistivity anomaly due to phase transition starts to broaden and the value of residual resistivity increases as is expected for an antiferromagnetic system.

          The low temperature electrical resistivity data is described well by the expression \cite{13},
\begin{displaymath}
\rho(T) = \rho_{0} + C \bigg\{ \frac{1}{5} T^5 + \triangle  T^4 + \frac{5}{3} \triangle ^2 T^3 \bigg\} exp(-\triangle /T)
\end{displaymath}
where $\rho_{0}$ is the residual resistivity, C is the coefficient to the magnon contribution, and $\triangle$  is the magnon energy gap. The experimentally observed data is reproduced very accurately by the above expression in the temperature range (2 -- 7 K) with  $\rho_{0}$ = 1.62 $\mu\Omega$-cm, C = 0.00017 $\mu\Omega$-cm/K$^5$, and the energy gap $\triangle$  = 3.40 K. This is in accordance with the antiferromagnetic magnon gap-like feature observed in the specific heat data.

           The lower inset of figure 4 shows the field dependence of resistivity at 2 K and 15 K plotted as magnetoresistance ($\Delta$$\rho$/$\rho$(0) = [$\rho$(B)-$\rho$(0)]/$\rho$(0), where $\rho$(B) is the resistivity measured at a magnetic field B). A linear field dependence of resistivity is inferred both in paramagnetic and antiferromagnetic states. The magnetoresistance is unusually high and positive in the ordered state. We observe almost 10 times increase in resistivity at a field of 10 T at 2 K, which leads to a magnetoresistance of $\sim$ 940\%, which is surprisingly large. Even in the paramagnetic state a large positive magnetoresistance is observed ($\sim$190\% at 10 T at 15 K).  Such a large positive magnetoresistance in an antiferromagnetically ordered polycrystalline intermetallic system is quite unusual and intriguing. Since isothermal magnetization in the ordered state shows metamagnetic transition at 1.6 T, one would expect a negative magnetoresistance in the vicinity of that characteristic field. However no such effect is observed in transverse magnetoresistance (magnetic field perpendicular to current). This might be due to the fact that large part of the resistance change is due to modification of the trajectory of the current carrying electrons rather than changes in the electronic density of states (closing of superzone gap etc.) by the application of magnetic field.

           A large positive magnetoresistance (as high as 85\%) has also been observed in antiferromagnetically ordered system R$_{2}$Ni$_{3}$Si$_{5}$ (R= Tb, Sm, Nd) \cite{14} where it was suggested that such behaviour could be the result of layered structure of rare-earth atoms which provides the sheets of ordered spins in analogy with that in Dy/Sc superlattice \cite{15}. A similar favourable situation also exists in the case of our compound Pr$_{2}$Pd$_{3}$Ge$_{5}$ which consists of layers of Pr-atoms separated by Pd-Ge network in \textit{Ibam}-structure. Therefore as in the case of R$_{2}$Ni$_{3}$Si$_{5}$ (R= Tb, Sm, Nd), the large positive GMR in Pr$_{2}$Pd$_{3}$Ge$_{5}$ may be speculated to arise from the layered nature of structure. We wish to emphasize here that the magnitude of the magnetoresistance is exceptionally large and might have strong technological relevance. Such a high value (in excess of 200 \%) of magnetoresistance is observed in Ballistic Ni Nanocontacts \cite{16}. Extremely large value of magnetoresistance (resistance decreases by a factor of 200) was also found in HgCr$_{2}$Se$_{4}$ which is caused by magnetic field induced changes in the carrier mobility and concentration \cite{17}.

           Apart from the unusually large magnetoresistance, the sharp drop of magnetization below 7.5 K, the sharp increase of magnetization above the critical field and the gapped behavior in the specific heat and resistivity are strong indication of a strongly anisotropic Ising-type antiferromagnetic system. Investigations on single crystals are desired to further understand the anisotropic magnetic properties of this compound.

\section*{B. Pr$_{2}$Rh$_{3}$Ge$_{5}$}

           The magnetic susceptibility data of Pr$_{2}$Rh$_{3}$Ge$_{5}$ at a field of 0.1 T is plotted in figure 5. The susceptibility curve is found to obey the modified Curie-Weiss behavior without any magnetic ordering down to 2 K. From a fit to the susceptibility data above 7 K with $\chi$ = $\chi_{0}$ + C/(T-$\theta_{p}$) we obtained $\chi_{0}$ = 1.18 x10$^{-3}$ emu/Pr-mole and the paramagnetic Curie-Weiss temperature $\theta_{p}$ = -11.5 K. The effective moment $\mu_{eff}$ came out to be 3.51 $\mu_{B}$, which is very close to the theoretical value of 3.58 $\mu_{B}$ for Pr$^{3+}$. The isothermal magnetization data at 2 K (shown in the inset of figure 5) show almost linear field dependence which is a characteristic of paramagnetic system, and the magnetization does not reach to saturation value up to 5 T field.

\begin{figure}
\includegraphics[width=8.5cm,keepaspectratio]{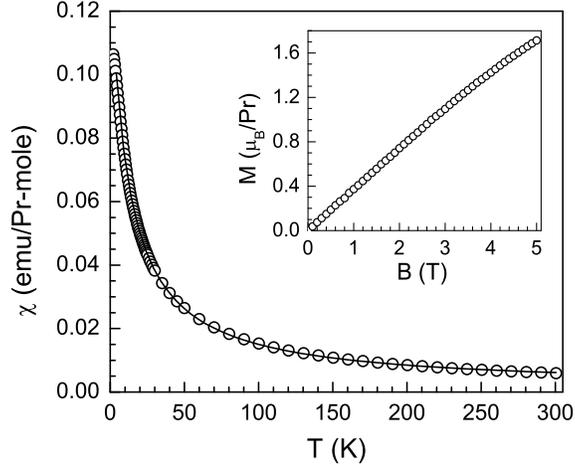}
\caption{\label{fig5} Temperature dependence of magnetic susceptibility of Pr$_{2}$Rh$_{3}$Ge$_{5}$ at 0.1 T. Solid line represents the fit to the modified Curie-Weiss behaviour.  Inset shows the field dependence of isothermal magnetization at 2 K.}
\end{figure}

\begin{figure}
\includegraphics[width=8.5cm,keepaspectratio]{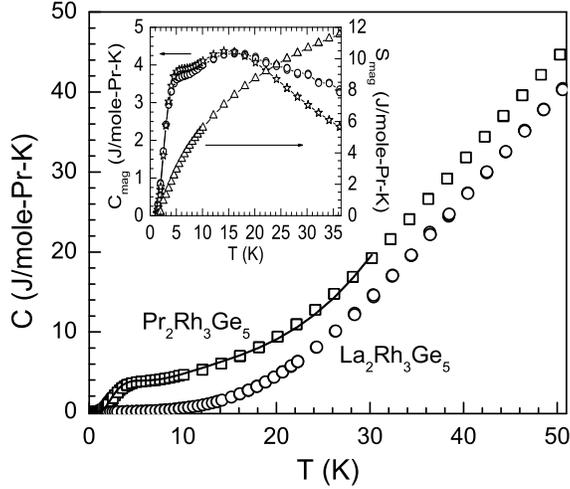}
\caption{\label{fig6} Temperature dependence of the specific heat of La$_{2}$Rh$_{3}$Ge$_{5}$ and Pr$_{2}$Rh$_{3}$Ge$_{5}$ in the temperature range 0.5 K -- 50 K. The solid line is the fit to the CEF scheme as described in the text. Inset shows the magnetic part of the specific heat and entropy as a function of temperature. The data shown by stars in the inset represent the magnetic part to the fit in CEF analysis.}
\end{figure}

           The specific heat data of La$_{2}$Rh$_{3}$Ge$_{5}$ and Pr$_{2}$Rh$_{3}$Ge$_{5}$ are plotted in figure 6. No anomaly is observed in the specific heat data of La$_{2}$Rh$_{3}$Ge$_{5}$ down to 2 K, and the $\gamma$ value is estimated to be $\sim$ 10 mJ/mole-La-K$^{2}$. In the case of Pr$_{2}$Rh$_{3}$Ge$_{5}$ also the specific heat data do not exhibit any pronounced anomaly due to a phase transition. We, however, notice a broad Schottky-type peak centered around 4.5 K. The solid line in figure 6 represents the crystal field analysis and include a phonon contribution equal to that of La$_{2}$Rh$_{3}$Ge$_{5}$ and an additional electronic contribution to take care for the difference in  Sommerfeld coefficients of La$_{2}$Rh$_{3}$Ge$_{5}$ and Pr$_{2}$Rh$_{3}$Ge$_{5}$. The magnetic contribution to the specific heat of Pr$_{2}$Rh$_{3}$Ge$_{5}$ which was estimated as the difference in the specific heats of Pr$_{2}$Rh$_{3}$Ge$_{5}$ and La$_{2}$Rh$_{3}$Ge$_{5}$ is shown in the inset of figure 6 together with the magnetic entropy and the crystal field analysis. The experimentally observed magnetic specific heat data at low temperatures are fairly reproduced by the four low lying crystal electric field levels (four CEF singlets at 0 K, 12 K, 40 K and 60 K) with a significant departure above 20 K due to the contribution from higher excited states. That the ground state is a singlet separated by 12 K from the first excited is further supported from the fact that magnetic entropy attains a value of Rln(2) at 10 K (inset of figure 6). The temperature dependence of magnetic entropy also reveals the significant contribution from higher excited states, e.g. an increased value of magnetic entropy over Rln(3) at 40 K in the magnetic entropy is due to the contribution from the levels above 40 K. Further a constant difference in the specific heats of Pr$_{2}$Rh$_{3}$Ge$_{5}$ and La$_{2}$Rh$_{3}$Ge$_{5}$, which starts appearing above 25 K, is an indication of the involvement of closely spaced excited states.  From such a behaviour of the specific heat we can suggest that in this case possibly the crystal field removes the nine fold degeneracy of Pr$^{3+}$ by splitting them into nine singlets which is expected in the orthorhombic environment. This CEF split nonmagnetic singlet ground state leads to no magnetic ordering in Pr$_{2}$Rh$_{3}$Ge$_{5}$. In the case of homologous compound Pr$_{2}$Rh$_{3}$Si$_{5}$ too after an analysis of crystalline electric field Ramakrishnan et al. have suggested a singlet CEF ground state [12].  The Sommerfeld coefficient $\gamma$ in Pr$_{2}$Rh$_{3}$Ge$_{5}$ is enhanced and has a value of $\sim$ 81 mJ/mole-Pr-K$^{2}$.

\begin{figure}
\includegraphics[width=8.5cm,keepaspectratio]{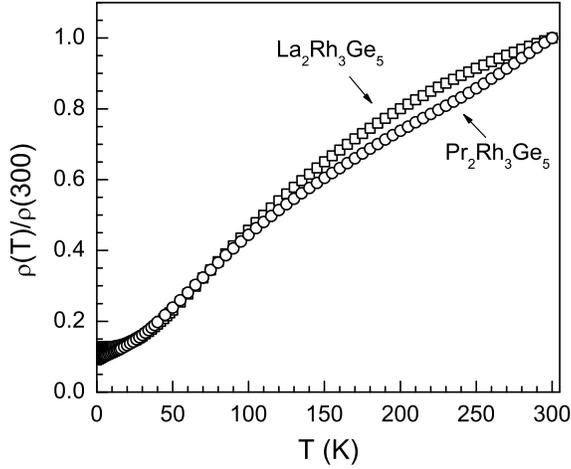}
\caption{\label{fig7} Normalized electrical resistivity of La$_{2}$Rh$_{3}$Ge$_{5}$ and Pr$_{2}$Rh$_{3}$Ge$_{5}$ as a function of temperature in the temperature range of 2 K -- 300 K.}
\end{figure}

          Figure 7 shows the electrical resistivity data of La$_{2}$Rh$_{3}$Ge$_{5}$ and Pr$_{2}$Rh$_{3}$Ge$_{5}$. The resistivity data of La$_{2}$Rh$_{3}$Ge$_{5}$ exhibit a very broad curvature around 150 K possibly due to the band effects together with a residual resistivity of $\sim$ 39 $\mu\Omega$-cm (at 2 K) and a residual resistivity ratio of 9. A considerable departure from the linear temperature dependence is also observed in resistivity of Pr$_{2}$Rh$_{3}$Ge$_{5}$ on account of crystal field effects and/or band structure effects. The value of residual resistivity at 2 K is $\sim$ 58 $\mu\Omega$-cm with the residual resistivity ratio of 11. The large residual resistivity is possibly due to the presence of (micro)cracks in the sample. In spite of the large residual resistivity in the sample we made an attempt to extract the physics content in the data. The data at low temperature (below 4 K) fits to $\rho$ (T) = $\rho_{0}$ + AT$^{2}$ with A = 0.25874 $\mu\Omega$-cm/K$^{2}$. This corresponds to A/$\gamma^{2}$ = 3.94 x 10$^{-5}$ $\Omega$-cm-K$^{2}$-mole$^{2}$-J$^{-2}$ which is larger than the universal value of Kadowaki-Woods ratio expected for the Ce-based Kondo lattice/heavy fermion compounds \cite{18}.

          Further, the Wilson ratio R$_{W}$ = 23.31 using the value of $\chi$ = 0.10641 emu/Pr-mole at 2 K (B = 0.1T) and $\gamma$ = 81 mJ/mole-Pr-K$^{2}$ is also large similar to the case of Non-Fermi-liquid heavy fermion compound YbRh$_{2}$Si$_{2}$ which is situated very close to the quantum critical point in which large Wilson ratio is attributed to the presence of ferromagnetic fluctuations \cite{19}. Since the results obtained for Pr$_{2}$Rh$_{3}$Ge$_{5}$ are in the same range of the values which are obtained for Ce- and Yb-based heavy fermion systems, Pr$_{2}$Rh$_{3}$Ge$_{5}$ is a new heavy fermion system.

          Since the electrical resistivity and specific heat data of Pr$_{2}$Rh$_{3}$Ge$_{5}$ do not show any signature of Kondo effect, the mechanism for the heavy fermion behavior in Pr$_{2}$Rh$_{3}$Ge$_{5}$ is not the usual Kondo effect, rather, the heavy fermion behavior results from the low lying crystal field levels by the inelastic scattering of conduction electrons with the excited levels usually referred as excitonic mass enhancement. The theory of excitonic mass enhancement was initially proposed by White and Fulde to explain the mass enhancement of Pr metal \cite{20,21} and has been recently applied to explain the origin of heavy fermion behaviour in  PrOs$_{4}$Sb$_{12}$ \cite{6}. There are two intrinsic requirements for the excitonic mass enhancement -- the ground state must be a singlet and the splitting energy between the ground state and the first excited state must be low -- to develop a heavy fermion state. For our system Pr$_{2}$Rh$_{3}$Ge$_{5}$ both the requirements are fulfilled -- the ground state is nonmagnetic singlet and first excited state lies at 12 K -- hence Pr$_{2}$Rh$_{3}$Ge$_{5}$ is a new heavy fermion system that validates the theory of excitonic mass enhancement.

          The magnetic and transport properties discussed above were reproduced in the second sample of Pr$_{2}$Pd$_{3}$Ge$_{5}$ and Pr$_{2}$Rh$_{3}$Ge$_{5}$ which establishes the intrinsic nature of the behaviour.

\section*{Summary and conclusion}

          We synthesized and investigated two new ternary compounds Pr$_{2}$Pd$_{3}$Ge$_{5}$ and Pr$_{2}$Rh$_{3}$Ge$_{5}$. In the case of Pr$_{2}$Pd$_{3}$Ge$_{5}$ the crystal field split ground state is doublet and the magnetization, electrical resistivity and specific heat clearly establish that the compound orders antiferromagnetically below 8.3 K. It also shows unusually large magnetoresistance. Pr$_{2}$Rh$_{3}$Ge$_{5}$, on the other hand, has a singlet ground state and does not show any kind of order down to 0.5 K. A large value of Sommerfeld coefficient (81 mJ/mol K$^{2}$) is found in this compound. Enhanced Wilson ratio, enhanced value of coefficient A together with large $\gamma$ establish Pr$_{2}$Rh$_{3}$Ge$_{5}$ as a possible candidate for electronic mass enhancement due to low lying crystal field excitations. Role of low lying crystal field excitation for this enhanced Sommerfeld coefficient should be explored further by direct experimental verification of the crystal field level scheme and low lying excitations using inelastic neutron scattering.

%add ref
%\bibliography{ref3}% Produces the bibliography via BibTeX.

\end{document}